\documentclass[aps,prl,twocolumn,showpacs,preprintnumbers,amsmath,amssymb]{revtex4}

\usepackage{graphicx}
\usepackage{dcolumn}
\usepackage{bm}

\begin{document}


\title{Collisional losses, decoherence, and frequency shifts \\
 in optical lattice clocks with bosons}%

\author{Ch. Lisdat,}\email{christian.lisdat@ptb.de}
\author{J.S.R. Vellore Winfred, T. Middelmann, F. Riehle, and U. Sterr}

\affiliation{Physikalisch-Technische Bundesanstalt, Bundesallee 100, 38116 Braunschweig, Germany}
%

\date{\today}

\begin{abstract}
We have quantified collisional losses, decoherence and the collision shift in a one-dimensional optical lattice clock with bosonic $^{88}$Sr. The lattice clock is referenced to the highly forbidden transition $^1$S$_0 - ^3$P$_0$ at 698~nm, which becomes weakly allowed due to state mixing in a homogeneous magnetic field. We were able to quantify three decoherence coefficients, which are due to dephasing collisions, inelastic collisions between atoms in the upper and lower clock state, and atoms in the upper clock state only. Based on the measured coefficients, we determine the operation parameters at which a 1D-lattice clock with $^{88}$Sr shows no degradation due to collisions on the relative accuracy level of 10$^{-16}$.
\end{abstract}

\pacs{06.30.Ft, 32.70.Jz, 34.50.Cx, 37.10.Jk}

\preprint{2.6}
\maketitle
%
%
%
%
%
%
%
Laser spectroscopy of transitions with millihertz linewidth has become the essential tool for the rapidly evolving optical clocks based on single ions \cite{ros08} or neutral atom clouds \cite{cam08b}. These clocks promise a realization of the base unit \lq second\rq~with higher stability and accuracy than the currently employed frequency standards using a microwave transition in $^{133}$Cs. Optical clocks furthermore enable tests of new theories, which predict e.g. the variation of fundamental constants \cite{lea07}, or they can be used for relativistic geodesy field experiments \cite{mue08c}. Due to long coherence times and very sensitive interrogation, optical clocks with neutral atoms are also ideally suited to explore the physics of ultra-cold collisions. The interpretation of collision experiments provides insight into the coupling between the atoms in different electronic states at ultra-low temperature. Understanding collision properties of alkaline earth atoms is required for proposed quantum computing schemes \cite{gor09} and to test atomic structure calculations \cite{mit04}. 

Strontium is widely used for magic wavelength optical lattice clocks with, both, bosons and fermions \cite{cam08b,bai08,aka08,bai07}. At ultra low temperature, collisions of fermions are strongly suppressed by the Pauli exclusion principle \cite{cam09}. The bosonic isotope $^{88}$Sr offers higher natural abundance and easier laser cooling. It is therefore of interest especially for transportable setups. Collisions are suppressed if a three dimensional lattice with an occupancy of at most one atom per site \cite{aka08} is used but the complexity of setup and experimental sequence is increased. The use of $^{88}$Sr in a 1D-lattice clock can therefore be advantageous if the effects of collisions were well quantified \cite{bai07, tra08}. 

Three effects induced by collisions are expected: inelastic losses, collision broadening of the transition, and a collision-induced frequency shift. We have quantified all three in our lattice clock setup. With our results we are able to predict operation conditions of a $^{88}$Sr lattice clock under which collisions are not expected to limit the relative accuracy on the level of 10$^{-16}$. Additional shifts due to clock-laser induced ac Stark effect and quadratic Zeeman effect in bosonic lattice clocks are theoretically well understood and can already be controlled with high precision. For these, a relative uncertainty far better than $10^{-16}$ is expected for the near future \cite{pol08}.

Our experiment was described before \cite{leg08}. In short, we first prepare the $^{88}$Sr atoms in a magneto-optical trap (MOT) operated on the strong 461~nm transition $^1$S$_0 - {^1}$P$_1$. The atoms are loaded for up to 250~ms from a Zeeman-slowed atomic beam, which is deflected into the MOT region by means of an optical molasses.

The atoms are further cooled in a second cooling phase on the 689~nm intercombination line $^1$S$_0 - {^3}$P$_1$. First, the cooling light is frequency modulated to enlarge the capture velocity range of the second stage MOT. Then, after 70~ms, the modulation is turned off and within another 70~ms the atoms are further cooled to below 4~$\mu$K. A horizontal 1D-optical lattice is overlapped during the whole cooling cycle with the MOT. It is created by a laser beam of up to 300~mW power at the magic wavelength of 813~nm. The beam is focused to a waist of 34~$\mu$m and retro-reflected into itself. The waist size is confirmed via measurements of trap oscillation frequencies.
\begin{figure}[t]
\footnotesize
\includegraphics[width=7.5cm]{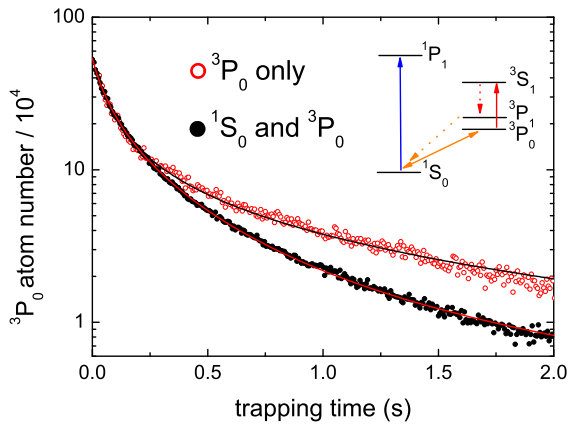}
\caption{(color online) Inelastic losses by collisions between $^3$P$_0$ atoms (open circles) and a mixture of $^3$P$_0$ and $^1$S$_0$ atoms (full symbols). Points are the measured data, lines are the results of fits. The inset shows schematically the involved atomic levels. The initial number of ground state atoms is about $3 \times 10^6$.}
\label{fig:losses}
\end{figure}
After the second cooling phase, up to $3 \times 10^6$ atoms are trapped in the optical lattice (about 1000 atoms per site). The atom number and temperature were determined by absorption image measurements after a variable expansion time. Untrapped atoms are falling out of the interrogation area within 150~ms. Then the trapped atoms are interrogated on the clock transition $^1$S$_0 - ^3$P$_0$ at 698~nm. The clock laser beam of 2.5~mW maximum power is overlapped with the lattice through a dichroic mirror and focused to a waist radius of about 39~$\mu$m at the position of the atoms. To induce a dipole transition matrix element on the clock transition, we apply a homogeneous magnetic field of up to 3~mT. 

After excitation on the clock transition the remaining ground state atoms $^1$S$_0$ are detected by recording the fluorescence during a 461~nm MOT phase and then pushed away by resonant 461~nm radiation. In the same way, excited $^3$P$_0$ atoms are detected after optically pumping them to the $^3$S$_1$ state from where they decay to the ground state via the $^3$P$_1$ state (inset Fig.~\ref{fig:losses}). Photoassociative losses by the repumping light are avoided by switching off the lattice and expanding the cloud of $^3$P$_0$ atoms before repumping \cite{tra08}. 

We first address {\it inelastic collisions}: Without excitation on the clock transition, we observe a background gas limited lifetime of the atoms in the lattice of 7~s to 8~s. In the presence of atoms in the $^3$P$_0$ state, density dependent shortening of the trap lifetime is observed. Inelastic two-body losses are possible by collisions of $^3$P$_0$ atoms with either $^3$P$_0$ or $^1$S$_0$ atoms. The decay can be modeled by the coupled differential equations
\begin{eqnarray}
\label{eq:dec}
\dot{\rho}_{\rm e} &=& -\Gamma \rho_{\rm e} -\gamma_{\rm ge} \rho_{\rm g} \rho_{\rm e} -\gamma_{\rm ee} \rho^2_{\rm e} \nonumber \\
\dot{\rho}_{\rm g} &=& -\Gamma \rho_{\rm g} -\gamma_{\rm ge} \rho_{\rm e} \rho_{\rm g}.
\end{eqnarray}
Here, $\rho_{\rm g}$ ($\rho_{\rm e}$) denotes the local atomic density in the ground (excited) state, $\Gamma$ is the inverse trap lifetime and $\gamma_{\rm ge}$  ($\gamma_{\rm ee}$) the loss coefficient for ground -- excited (excited -- excited) state collisions. Radiative decay is much slower and neglected.

To distinguish experimentally between both inelastic collision channels, we first removed the ground state atoms from the lattice and observed the decay of the $^3$P$_0$ population only (Fig.~\ref{fig:losses}). In this case, the Eqs.~\ref{eq:dec} simplify, decouple, and can be solved analytically. Integration over the spatial coordinates of a single lattice site provides an expression for its atom number:
\begin{equation}
\label{eq:decee}
N(t) = N_0 \frac{\exp{(-\Gamma t)}}{1 + N_0 \gamma_{\rm ee}/ (\pi^{3/2} \Gamma w^2_{\rm r} w_{\rm z}) \left[ 1-\exp(-\Gamma t)\right] }.
\end{equation}
$N_0$ denotes the initial atom number in a lattice site, $w_{\rm r}$ and  $w_{\rm z}$ are the $1/{\rm e}^2$-radii of the atomic cloud in a lattice site at a given temperature and set of lattice parameters.

In our experiment many independent lattice sites are occupied with different atom numbers. We numerically sum Eq.~\ref{eq:decee} over the lattice sites. The initial axial population distribution is described by a Gaussian of 1/e$^2$ radius $w'$ (determined from absorption images). From several data sets we determined the loss coefficient $\gamma_{\rm ee} = (2.0 \pm 0.7)\times 10^{-18}$~m$^3$/s. This is in good agreement with the value given by Traverso et al. \cite{tra08}. Our uncertainty is dominated by the uncertainty of the calculation of the radii $w_i$ and the scatter of the fit parameters. We neglect evaporation losses, since the lattice depth of $k_{\rm B} \cdot 45~\mu$K is much larger than the atomic thermal energy of below $k_{\rm B} \cdot 4~\mu$K.

We then investigated a mixed sample of $^3$P$_0$ and $^1$S$_0$ atoms. Collisions $^3$P$_0 + ^1$S$_0$ are responsible for the different decays visible in Fig.~\ref{fig:losses}.  We determined the loss coefficient  $\gamma_{\rm ge} = (2.3 \pm 1.1) \times 10^{-19}$~m$^3$/s by numerical integration of Eqs.~\ref{eq:dec}. Contributions to the uncertainty are again the determination of the $w_i$ and details of the fit. Careful consideration of the fitting strategy is required because losses are dominated by $^3$P$_0 + ^3$P$_0$ collisions as analyzed above. Best accordance of data and fit was obtained by fitting $\gamma_{\rm ee}$ together with $\gamma_{\rm ge}$. Nevertheless, we constrained $\gamma_{\rm ee}$ to the value obtained under the clean conditions of the first set of measurements. We also found slightly different results for $\gamma_{\rm ge}$ depending on whether we fitted only the excited state population, the ground state atom number or both simultaneously. We attribute this scatter to inelastic collisions that diminish the $^3$P$_0$ atom number during the detection of the ground state atoms. We corrected the excited state population for collisions with $^3$P$_0$ atoms but neglect other types of collisions. The scatter of $\gamma_{\rm ge}$ determined by different strategies is finally represented by the uncertainty given above.

Different loss mechanisms are possible at the respective collision asymptotes. The asymptote $^3$P$_0 + ^3$P$_0$ correlates at short internuclear distances to a $^1\Sigma^+_g$ state, which can decay by fluorescence to lower molecular states. More likely as source for the losses are non-adiabatic transitions to molecular potentials dissociating to lower asymptotes. In the case of collisions at $^1$S$_0 + ^3$P$_0$, the latter path seems to be very unlikely. Only one potential dissociates to the one lower asymptote $^1$S$_0 + ^1$S$_0$ and it is very steep at the energy of the $^1$S$_0 + ^3$P$_0$ asymptote. A direct spontaneous decay to that ground state potential is not possible due to selection rules but at short internuclear distance complex coupling exists to molecular states that are subject to spontaneous decay \cite{all05}. This more indirect pathway can be seen as a qualitative explanation for the significantly smaller loss coefficient at the asymptote $^1$S$_0 + ^3$P$_0$ compared to the asymptote $^3$P$_0 + ^3$P$_0$. However, we cannot exclude three-body losses or processes involving a lattice photon.

\begin{figure}[t]
\footnotesize
\includegraphics[width=8.5cm]{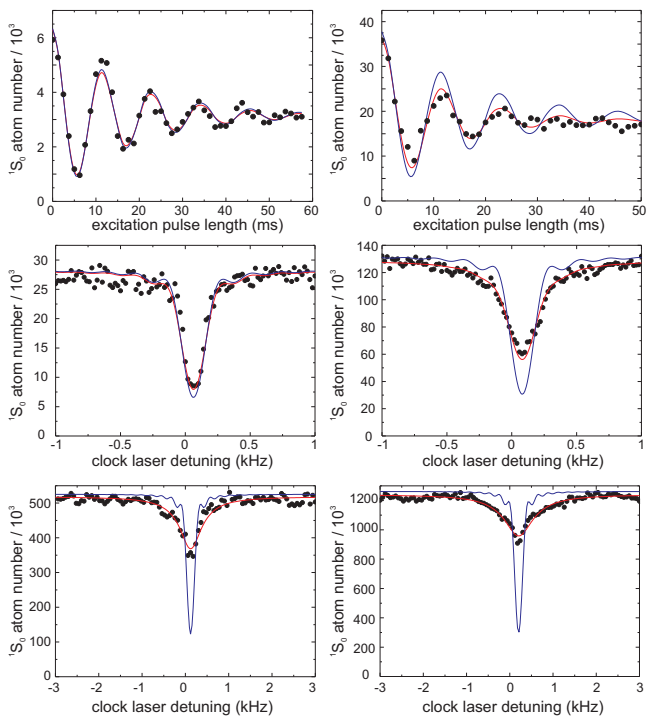}	
\caption{(color online) Rabi oscillations (top row) and spectra of the clock transition (below) for different atom numbers indicated on the ordinate. The solid lines are calculated with the density matrix model. The red curves stem from fits of the dephasing constant $\gamma_{\rm dep}$, while the blue ones are a calculations with $\gamma_{\rm dep} \equiv 0$.}
\label{fig:broad}
\end{figure}

As second decoherence process in a $^{88}$Sr lattice clock we identified strong {\it density dependent broadening} of the clock transition (see Fig.~\ref{fig:broad}). To model the observed spectra we extended the differential equations Eqs.~\ref{eq:dec} to a formalism of the full density matrix $\rho$ ($\rho_{11}=\rho_{\rm g}$, $\rho_{22}=\rho_{\rm e}$) including the excitation dynamics, and integrated them numerically:
\begin{equation}
\label{eq:rho}
\dot{\rho} = -\frac{i}{\hbar}[H,\rho] + \mathcal{R}(\rho).
\end{equation}
In the rotating wave approximation, the atom-light Hamiltonian and the relaxation matrix $\mathcal{R}(\rho)$ become
\begin{eqnarray}
\label{eq:ham}
\frac{H}{\hbar} &=& \left( \begin{array}{cc}
    		0	& \Omega/2  \\
\Omega/2	& \Delta    \\
\end{array} \right) \nonumber \\
\mathcal{R}(\rho)_{11} &=& -(\Gamma + \gamma_{\rm ge}\rho_{\rm 22})\rho_{11}	\nonumber \\
\mathcal{R}(\rho)_{12} &=& \mathcal{R}(\rho)_{21}^* = -[\gamma_{\rm ge}(\rho_{\rm 11}+\rho_{\rm 22})/2 + \gamma_{\rm ee}\rho_{\rm 22}/2 \nonumber \\
&& + \Gamma + L + \gamma_{\rm dep}\rho_{\rm 11} ] \rho_{12}  \nonumber \\
\mathcal{R}(\rho)_{22} &=&  -(\Gamma + \gamma_{\rm ge}\rho_{\rm 11} + \gamma_{\rm ee}\rho_{\rm 22})\rho_{22}.
\end{eqnarray}
The Rabi frequency $\Omega$ can be calculated from the experimental parameters \cite{tai06,bai07}, the detuning $\Delta$ is determined from the spectrum itself. Dephasing or damping of the coherences in the density matrix is introduced by the coefficient $L$ representing an effective laser linewidth and inhomogeneous damping due to spatial dependent Rabi frequencies. The dephasing by elastic collisions is described by $\gamma_{\rm dep}$. We assume dephasing by collisions with ground state atoms. Different mechanisms can be imagined but the experimental data do not allow to distinguish between them. 

For the sake of fast computation, we neglected spatial variations in the excitation probability due to the lattice or the clock laser profile. Furthermore, a rectangular population distribution of full width $2w'$ in axial direction was assumed. We have verified in the case of the inelastic losses that the effect of this approximation is minor. 

To fix the excitation dynamics we fitted observed Rabi oscillations and a scan of the clock transition at an atom density, which does not significantly broaden the transition. The fitted Rabi frequency is about 17\% smaller than the one calculated from the peak clock laser intensity and the homogeneous magnetic field, an effect we attribute to the finite Lamb-Dicke  parameter. The FWHM laser linewidth was fitted to be about 36~Hz, which is roughly consistent with a transition linewidth of 35~Hz we have observed at an excitation pulse length of 35~ms and a few thousand atoms. To match the observations at higher density we had to include (besides the inelastic collisions determined before) a dephasing coefficient of $\gamma_{\rm dep} = (3.2 \pm 1.0 ) \times 10^{-16}$~m$^3$/s (see Fig.~\ref{fig:broad}). The effect is stronger by a larger factor (10$^3$) compared to collisions $^1$S$_0 + ^3$P$_1$ in a 1000~K Sr gas \cite{cra94}, which could indicate the existence of a scattering resonance. We did not observe a significant temperature dependence of any collision coefficient $\gamma$ between 2~$\mu$K and 4~$\mu$K.

The third collision effect being very important for the accuracy of an optical clock with neutral atoms is the {\it frequency shift} $\delta \nu$ due to collisions. We have measured $\delta \nu$ by two stabilizations with interleaved probing cycles \cite{deg04, deg05a}. The clock laser probes either about $10^4$ atoms (used as reference) or a higher atom number (and thus density). This is systematically varied via the duration of the initial MOT stage. The density shift is revealed as difference of the offset frequencies for the alternated stabilizations between the clock laser and its reference cavity. We have verified that no systematic frequency shifts occur for equal atom numbers on the level of a few 10~mHz. We shortened the cycle time to $\tau_{\rm cycle} = 200$~ms and detected the excited state population only. This way the stability of the difference relative to the optical frequency improved to be better than $10^{-15}$ at an averaging time $\tau = 10$~s and scaled as $1/\sqrt{\tau}$.
\begin{figure}[t]
\footnotesize
\includegraphics[width=7.5cm]{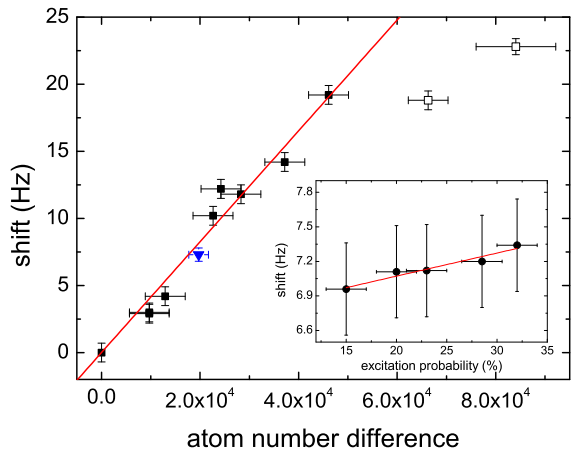}	
\caption{(color online) Observed density dependent shift of the $^{88}$Sr clock transition. The open symbols were excluded from the linear fit (full line, see text). The inset shows the dependence of the shift on the excitation probability with a linear fit. The triangle in the main graph represents the data in the inset. The shift was rescaled according to the trap parameters and atom temperature.}
\label{fig:shift}
\end{figure}
The dependence of the observed frequency shift as a function of the difference of atom number in the two stabilizations is shown in Fig.~\ref{fig:shift}. At low density we observe a linear dependence of the shift on the atom density; for high density we observe saturation of the shift. We attribute this to the changed dynamics during the excitation due to losses and dephasing. A theoretical analysis of the excitation process including a excitation dependent frequency shift will be carried out in the future. The clock laser was stabilized by probing the atoms with a 5~ms $\pi$-pulse detuned such that the excitation probability was reduced to about 35\% after the pulse.

From a linear fit through the origin (excluding the open data points in Fig.~\ref{fig:shift} because of saturation), the coefficient of the absolute collision shift as a function of the atom number $N$ can be determined to better than 4\%. The coefficient for the shift as function of the density is $\Delta \nu_{\rm \rho} = 2 \,\delta\nu \, \pi^2 w' w_{\rm r}^2 w_{\rm z}/N \lambda = (7.2 \pm 2.0)\times 10^{-17}$~Hz$\cdot$m$^3$ with an uncertainty mainly due to the density determination ($\lambda$: magic wavelength). A temperature dependence of the shift was not observed between 2~$\mu$K and 4~$\mu$K. 

The inset in Fig.~\ref{fig:shift} shows the dependence of the density shift on the excitation probability after the clock laser pulse, which was varied by the clock laser detuning in the stabilization cycle. We observe a weak relative influence of $(2.9 \pm 4.5)\times 10^{-3}$ per percent excitation probability.

Having quantified three collision influences, we give guidelines for the design of a 1D-lattice clock with bosonic $^{88}$Sr. Assuming a typical lattice depth of $k_{\rm B} \cdot$10~$\mu$K, an atom temperature of 3~$\mu$K, and an available lattice laser power of 300~mW, one could choose a lattice waist of 75~$\mu$m. With $w' = 280$~$\mu$m and at the current level of accuracy for the density shift correction of 4\%, a density shift of about 1~Hz would be tolerable to reach a relative accuracy of $10^{-16}$. This would limit the atom number to about $2 \times 10^4$, a value comparable to or larger then in present lattice clocks with $^{87}$Sr. The collisional broadening is then about 1.3~Hz. A new density shift measurement in the proposed lattice should yield an improved correction and allows for increasing the atom number until the collisional broadening becomes relevant. Aiming at a line width of about 10~Hz, operation with more than $10^5$ atoms is feasible. With a cycle time of 200~ms, the quantum projection noise limited relative stability in one second is $2\times 10^{-17}$. At this density, according to our findings losses do not distort the observed line or limit the excitation probability.

We conclude that collisions do not impose limitations for a bosonic $^{88}$Sr lattice clock compared to state-of-the art ones with $^{87}$Sr. We expect that these newly observed and quantified decoherence mechanisms also will have impact on the design and operation of experiments with alkaline earth atoms that require long coherence times like quantum computing, millihertz linewidth lasers \cite{mei09}, or optical Feshbach resonances \cite{nai06}.

The support by the Deutsche Forschungsgemeinschaft in SFB~407, by the Centre of Quantum Engineering and Space-Time Research (QUEST), the European Community's ERA-NET-Plus Programme under Grant Agreement No.~217257, and by the ESA and DLR in the project Space Optical Clocks is gratefully acknowledged. 
%


\end{document}